\documentclass[11pt,a4paper]{samo}

\usepackage[]{graphicx}
\usepackage[square,numbers]{natbib}

\title{Ground Layer Adaptive Optics: PSF effects on ELT scales}

\author{C. Arcidiacono\supit{1} and R. Ragazzoni\supit{1}}

\affiliation{\supit{1}INAF Astrophysical Observatory of Arcetri, Largo Enrico Fermi 5, Firenze 50125, Italy; E-mail: carmelo@arcetri.astro.it \\
  }

\begin{document}
  \maketitle

%%%%%%%%%%%%%%%%%%%%%%%%%%%%%%%%%%%%%%%%%%%%%%%%%%%

\begin{abstract}
On certain extent the behavior of the Adaptive Optics correction for Extremely Large Telescope scales with diameter size.
But in Ground Layer Adaptive Optics the combined effect of a Large Field of View and the large overlap of Guide Stars pupil footprints at high atmospheric altitude introduces severe changes in the behavior of the correction returning a very different distribution of the energy going from known 8-10meter to 100m diameters. In this paper we identify the reasons and the ways of these different behaviors.
\end{abstract}

%>>>> Include a list of keywords after the abstract


%%%%%%%%%%%%%%%%%%%%%%%%%%%%%%%%%%%%%%%%%%%%%%%%%%%

\section{INTRODUCTION}
\label{sect:intro}  % \label{} allows reference to this section
The scientific reasonable for Extremely Large Telescopes (ELT) is one
of the most investigated issue within the ELT projects. Science cases influence the road-map for the
scientific instruments and telescopes design, as well. In the frequently asked questions (FAQ)
list of different ELT projects (OWL\cite{owl}, TMT\cite{tmt}, EURO50\cite{euro50}) one of the most common is about the shape of PSFs for different instrument cases.
Of course detailed PSF shapes can be computed if the characteristics of the telescope and of the Adaptive Optics (AO) are known.
To date AO is far to be finally designed, even if there is a general consensus about the AO need in ELT telescope design
to take advantage of the improved resolution given by the extremely large diameter (D).

\section{From 10 to 100 meters diameter}
\label{sect:1}
\begin{figure}[h]
   \begin{center}
   \begin{tabular}{c}
   \includegraphics[width=16cm]{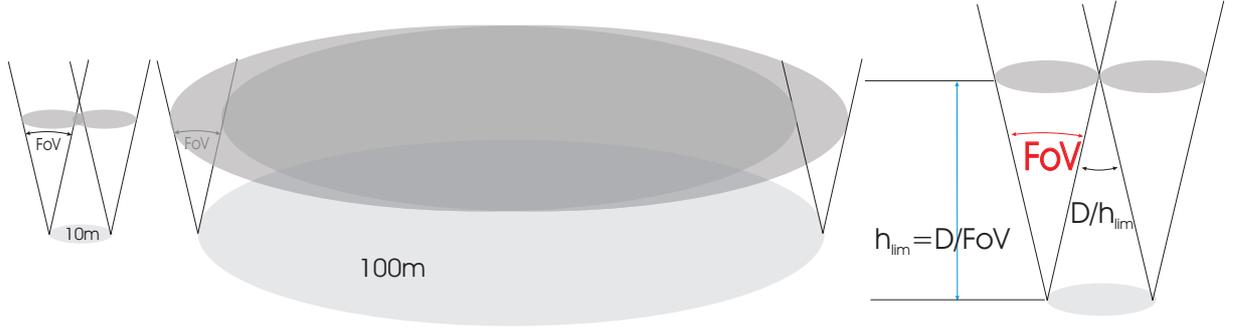}
   \end{tabular}
   \end{center}
   \caption[example]
%>>>> use \label inside caption to get Fig. number with \ref{}
   { \label{fig:1} In this figure are graphically presented the components of the phase aberration in the WF observed by 10m (left) and 100m telescope (right), after the GLAO correction. In the 10 meter case all the turbulence relative to high altitude layers introduce WF-aberrations in the whole spatial frequency domain, while in the 100m star WF the all low order modes (high layers too) are corrected by the GLAO system. In both cases the Ground Layer turbulence has been completely corrected. On the right the picture the h$_{\rm lim}$ is explained: it represents a typical altitude or a characteristics altitude where the layer phase aberration cannot be reconstructed using stars at the limit of the technical FoV used for WFS measurements.}

   \end{figure}

Going from 10 to 100m scales the range of spatial frequencies reachable by the system increases of a factor 10 for both $x$ and $y$ direction of the plane ($u$ and $v$ in the Fourier space). These new frequencies allows to obtain $\lambda/100$ resolution performance (it is $\propto$D$^{-1}$). But in the scaling operation not only the pupil of the telescope is involved, all the geometrical proportions scale with D. Similarly for an AO system we know that to achieve the same degree of correction we need to sample the wave--front (WF) aberrations with the same degree of accuracy. Technically speaking a 8$\times$8 sampling of the pupil for a 10m telescope corresponds to a 80$\times$80 on a 100m (one sub-aperture/r$_0$ @NIR band) and, respectively, scales in the same way the number of actuators needed to adequately compensate the WF $\approx 8^2$ to $\approx 80^2$. The basic dimensions of the AO systems seem to scale  $\propto$D$^2$. But, what does it happen to the conjugation altitude for the deformable mirrors (DM) or for AO-loop frequency? The configuration of these AO parameters will depend severely by the characteristics of the atmosphere, that does not scale with the telescope diameter: neither layer altitudes neither atmospheric parameters are changed because of the scaling. Considering the same corrected Field of View (FoV) and unchanged atmosphere a first immediate consequence is that the projected footprints of the telescope aperture along two different directions has a percentile overlap much larger in the 100m case than in the 10m. Figure~\ref{fig:1} presents two footprints for a 10m telescope and a 100m telescope at a given altitude and for a given identical FoV. The limiting altitude h$_{\rm lim}$ where footprints are totally separated and the turbulence is uncorrelated from one direction to the other, scales with D (h$_{\rm lim}(100) = 10\times$h$_{\rm lim}(10)$).
For 100m-telescope class cases in the atmospheric turbulence altitude range (0km-20km) the footprints largely overlap and the star WFs correlation at different direction is high. This is particularly true for the lower modes of correction that, hence, will be essentially identical for various points over the FoV (see Table~\ref{tab:1}).
\begin{table}[h]
\caption{In this table are listed the characteristic limiting altitude for different telescopes and different telescope diameters. For diameters larger than 30m the limiting altitude is much larger than the uppermost turbulent layers in the atmosphere.}
\label{tab:1}
\begin{center}
\begin{tabular}{lcccc} %% this creates four columns
%% lcccc => first column left justify, subsequent center justify
%% use of \rule[]{}{} below opens up each row

\hline
h$_{\rm{lim}}$ & 8m D& 30m D& 60m D& 100m D\\
\hline
2' FoV & 14km  & 51km & 103km & 171km \\
4' FoV &  7km  & 26km &  52km &  86km \\
6' FoV &  5km  & 17km &  34km &  57km \\
\hline
\end{tabular}
\end{center}
\end{table}
For D$<$10m cases the high layers are not sensed and WFs are uncorrelated on FoV$>2'$ (see Table~\ref{tab:1}).
We have to remind here that the ability to see and to correct not conjugated layers depends on the separation of stars (in other terms by the FoV used to sense and to correct): in fact non conjugated layers are seen more and more smoothed increasing the conjugation layer distance ($h$), blur$(\nu)$$\propto$FoV$\cdot$$\nu$$\cdot$$h$~\cite{mfov}.
According to this discussion GLAO systems performances cannot be scaled directly from 10 to 100m telescopes case, while MCAO systems can, as long as they are not based on the correction of one only layer but all layers.
Using GLAO the turbulence due to the higher layers cannot be seen nor corrected by a single mirror on the pupil plane in the case of the 10m class telescope, while in the 100m telescope case, the higher layers are sensed and corrected by the GLAO system, even if only partially because of the distance from the conjugation altitude (usually the ground layer).
In any case this implies a more efficient correction for a 100m class telescope.
\subsection{Outer-scale effect}
\label{sect:1.1}
\begin{figure}[h]
   \begin{center}
   \begin{tabular}{c}
   \includegraphics[height=2cm]{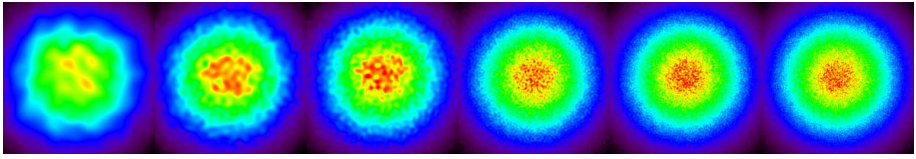}
   \end{tabular}
   \end{center}
   \caption[example]
   { \label{fig:2} Using LOST we simulate a 10 Layers atmosphere with good seeing condition ($\sigma_K=0.36''$) and L$_0$=25m. We used the same atmosphere in order to compute an integrate exposure for 6 different telescopes with 10, 20, 30, 60, 80 and 100m diameter respectively. The picture shows 6 open loop PSFs integrated for 5 seconds, from left to right increasing diameter. All PSFs have been normalized. }
   \end{figure}

\begin{figure}[h]
   \begin{center}
   \begin{tabular}{c}
   \includegraphics[height=5 cm]{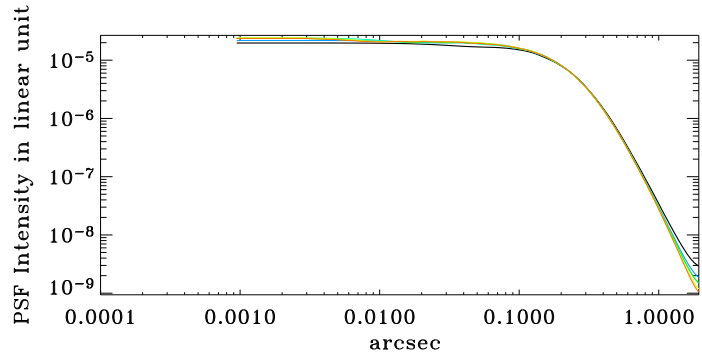}
   \end{tabular}
   \end{center}
   \caption[example]
   { \label{fig:3} This picture shows the seeing limited PSF intensity profile for 6 classes of telescope diameter (such as in Figure~\ref{fig:2}). No appreciable difference can be notice in the energy distribution and any peak due to large D/L$_0$ ratio can be underlined.}
   \end{figure}

The behavior of the adaptive correction is severely influenced by atmosphere statistical characteristics (such as r$_0$, $\tau_0$, outer-scale). In particular on diameter size several times larger than the turbulence L$_0$ starts to play a role also without any correction, in open loop conditions.
Current atmospheric measurements present outer-scales (L$_0$) values between 20m-100m, related to a
Von Karman power spectrum for the spatial distribution of the optical turbulence.
The diameter of the
telescope becomes comparable, or even larger than this outer scales value:
this implies that for ELT the tilt due to the atmospherical turbulence becomes sensitively small compared to the overall tilt experienced by 10m class telescope. Of course, the tilt due to the wind buffeting or a tracking error becomes predominant, but this one is full sky isoplanatic and it has typical frequencies much smaller than the optical one.
The statistics of the turbulence seen at ELTs scale is not Kolmogorov any more.
In fact, if the ratio D/L$_0$ increases up to infinity, the distribution of energy onto the spatial frequencies, $\nu$, is more and more flat,
and on $\nu$ relative to size larger than L$_0$ it shows a plateau.
A trivial effect is on the DM stroke needs because tip-tilt becomes more and more small, going in the direction favorable to the design of deformable mirrors.
In open loop small outer-scales behave such as a correction, moving the energy from the seeing halo disk to the diffraction peak. But this effect becomes important for outer-scales smaller than 15 meters\cite{conan}. We checked this statement running open loop simulation using the LOST\cite{lost} simulation tool and atmosphere with Von Karman power spectrum and L$_0$=25m, Figures~\ref{fig:2}\&\ref{fig:3}.
\section{GLAO PSF}
\begin{figure}[h]
   \begin{center}
   \begin{tabular}{c}
   \includegraphics[height=8cm]{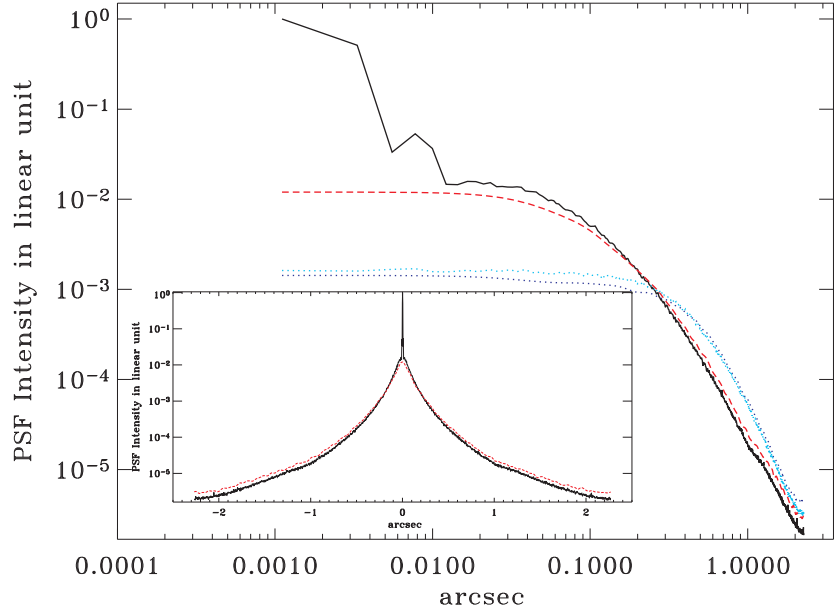}
   \end{tabular}
   \end{center}
   \caption[example]
   {\label{fig:5} This figure shows the expected PSF profiles for 10 (in dashed-red) and 100m telescopes. The plot is in log-scales, while the inset shows the two PSF profiles with arc second linearly plotted. The dotted lines represent the profile of the seeing disk for both 10 and 100 meter telescopes.}
   \end{figure}

GLAO is an AO system that aims to achieve large sky coverage, also at high galactic latitude, using very large FoV of the order of $4'-6'$ to look for reference stars and one DM conjugated to ground layer: this system corrects pretty well ground layer (theoretically GLAO could remove it completely), even if the thickness of the corrected atmosphere is small, scaling with FoV$^{-1}$.
We saw that up to 8m diameter h$_{\rm lim}$$<$7km is much lower than the maximum altitude of several turbulent layers (according to the model this altitude is H$_{\rm max}$=15-20km), then high layers ($>$7km) are unseen and uncorrected. This implies that, while ground layer is well corrected the high layers are not corrected at all, and on the corrected PSF they work generating a small seeing disk of the order of  $\lambda$/r$_{\rm 0,high}$ (r$_{\rm 0,high}$ is the coherence length of the high turbulence). Over 20m diameters the h$_{\rm lim}$ becomes higher than H$_{\rm max}$ and also high layers are low order corrected: on extra-large telescopes all possible layers are seen by the sensor and corrected by the DM, even if with decreasing accuracy for the higher ones. When h$_{\rm lim}$$>$H$_{\rm max}$, the PSF is composed by the diffraction limited spot over a depressed halo-seeing disk. ELT GLAO moves the energy in the central peak directly without an efficient high-seeing reduction, or energy concentration: it benefits of the ground removal and the high layer seeing disk defines the PSF energy distribution, excepting the diffraction-limited spike, which is not avoidable. Increasing the correction by using smaller sub-aperture for the pupil sampling, or using smaller technical FoV the Strehl Ratio (SR) is more and more high, while the seeing halo more and more depressed.
In a 10m GLAO system the higher layers are not sensed and then they cannot be corrected: the contribution of the higher layers on the total PSF is like a small seeing disk of $\lambda$/r$_{\rm 0,high}$. On the other hand, for a 100m telescope, the high layers are partially measured and corrected by the system and they contribute to the total PSF generating a depressed halo seeing disk with a small diffraction limited peak (Figure~\ref{fig:5}).
   \begin{figure}[h]
   \begin{center}
   \begin{tabular}{c}
   \includegraphics[width=8cm]{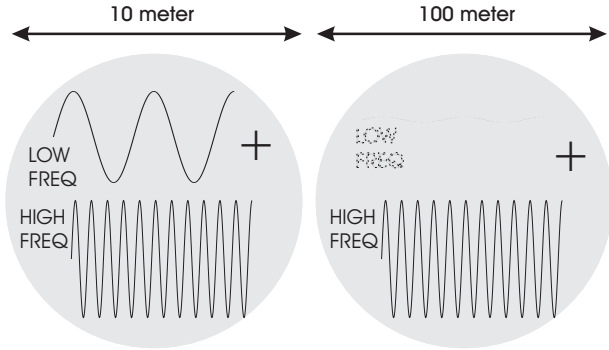}
   \end{tabular}
   \end{center}
   \caption[example]
   { \label{fig:4} In this figure the components of the phase aberration in the WF observed by 10m (left) and 100m telescope (right) after the GLAO correction are graphically presented. In the 10 meter case all the turbulence relative to high altitude layers introduces WF-aberrations in the whole spatial frequency domain, while in the 100m star WF the all low order modes (high layers too) are corrected by the GLAO system. In both cases the Ground Layer turbulence has been completely corrected.}
   \end{figure}

\section{Conclusions}
For a 100m telescope, D/r$_0$ is extremely large and the size of the halo ($\lambda /r_0$) compared to the core ($\lambda/D$) becomes huge, and then the surface brightness of the halo is very faint. We notice that a Strehl Ratio of 30\% or D=100m is equivalent to a Strehl Ratio of 99.5\% for D=8m. This would suggest that an AO system for an ELT can reach quite easily the Planer Finding regime thanks to the high contrast peak/halo. For ELTs, Ground Layer Adaptive Optics and MCAO are both possible to extend the isoplanatic patch, but the PSFs of both systems are quite different.
Many astronomical tasks (for example in extragalactic astronomy or cosmology) require more energy concentration than good narrow PSF for signal to noise reasons (background), requirements that go in the opposite direction: towards the correction of the high frequencies.

In this paper we have shown that Ground Layer AO on ELT could retrieve at most PSF large as  $\lambda$/r$_{\rm 0,high}$ that depends on the site (usually $\approx0.3''$ at K-band), but can be much narrow in very special site as DOME-C ($\approx 0.1''$) characterized by very small high layer turbulence. On this seeing reducted disk the diffraction limited peak due to the partial correction of the whole atmosphere is more and more important increasing the correction quality. Finally we stress that on point like source and not severely contrasted science target this low SR diffraction limited peak will give a great advantage.
%-------------

%%%%% REFERENCES  %%%%%

% You can either format the reference material directly or use BibTeX
\bibliographystyle{nature}
\bibliography{arcidiacono}
%\begin{thebibliography}{99}
%% the last item specifies width of reference number column

%\bibitem[1]{owl} Dierickx, P. et al. \spie {\bf 5489}  (2004), 391\\
%\bibitem[2]{tmt} {Sanders}, G. and {TMT Project} {\it Am. Astr. Soc. Meeting Abstracts} 205 (2004), 79\\
%\bibitem[3]{euro50} Andersen T.E. et al. \spie {\bf 5489}  (2004), 407\\
%\bibitem[4]{mfov} Ragazzoni, R. et al. {\it A\&A} 396 (2002), 731\\
%\bibitem[5]{conan} Conan, R. et al. {\it RevMexAA (Series de Conferencias)} 19 (2003), 31 \\
%\bibitem[6]{lost} Arcidiacono, C. et al. {\it Appl.Optics} {\bf 43} 22 (2004), 4288\\

%\bibitem[1]{owl} Dierickx, P., Brunetto, E.T.,  Comeron, F., Gilmozzi, R., Gont\'e, Fr\'ed\'eric Y. J., Koch, F., le Louarn, M., Monnet, G.~J., Spyromilio, J., Surdej, I., Verinaud, C., Yaitskova, N. \spie {\bf 5489}  (2004), 391\\
%\bibitem[2]{tmt} Nelson J.E.  \spie {\bf 5489}  (2004), 27.\\
%\bibitem[3]{euro50} Andersen T.E., Ardeberg A., Enmark A., Knutsson P., Morau D.,Owner-Petersen M., Riewaldt H., \spie {\bf 5489}  (2004), 24.\\
%\bibitem[4]{mfov} Ragazzoni, R., Diolaiti, E., Farinato, J., Fedrigo, E., Marchetti, E., Tordi, M., Kirkman, D. {\it A\&A} 396 (2002), 731\\
%\bibitem[5]{conan} Conan, R. et al. {\it RevMexAA (Series de Conferencias)} 19 (2003), 31-36 \\
%\bibitem[6]{lost} Arcidiacono C., Diolaiti E., Tordi M., Ragazzoni R., Farinato J., Vernet E. and Marchetti E. {\it Appl.Optics} 43 22 (2004), 4288\\

%\end{thebibliography}

\end{document}